\documentclass[prd,superscriptaddress,floatfix,showpacs,10pt,letterpaper]{revtex4}
\usepackage{graphicx} \usepackage{subfigure} \usepackage{amssymb}
\usepackage{verbatim} \usepackage{amsmath} \usepackage{amscd}
\usepackage{amssymb} 
\usepackage{epsfig}     

\input{epsf}
\usepackage{epsfig}
\usepackage{amssymb}
\usepackage{amsfonts}
\usepackage{amsbsy}
\usepackage{amsmath}

\usepackage{epstopdf}
\usepackage{graphics}      
\usepackage{graphicx}      

\newcommand{\spd}[1]{\textbf{d}#1}
\newcommand{\sph}{\star_0}

\begin{document}
\title{Fragmentation of black hole scaling throats in $\mathcal{N}=2$ supergravity}

\author{Hyeyoun Chung}  
\email{hyeyoun@physics.harvard.edu}
\affiliation{Jefferson Physical Laboratory, Harvard University,\\ 17 Oxford St., Cambridge, MA 02138, USA}
\date{\today}  
\begin{abstract}We find an instanton analogous to the Brill instanton that describes the fragmentation of a single-centered black hole scaling throat of charge $\Gamma_1 + \Gamma_2 + \Gamma_3$ in $\mathcal{N}=2$ supergravity to three disconnected throats of charges $\Gamma_1, \Gamma_2, \Gamma_3$, in the limit where the intersection products between the charges of the three throats satisfies $\langle \Gamma_1, \Gamma_2\rangle \lll \langle \Gamma_2, \Gamma_3\rangle, \langle \Gamma_3, \Gamma_1\rangle$. We evaluate the Euclidean action for this instanton and find that the amplitude for the tunneling process is proportional to the difference in entropy between the initial and final configurations.\end{abstract}
 \maketitle 

\section{Introduction}\label{sec-Intro}

In \cite{Brill}, Brill found an instanton that corresponds to the fragmentation of a single charged $AdS_2\times S_2$ universe into several disconnected $AdS_2\times S_2$ universes, and interpreted this instanton as describing topological fluctuations near the horizon of extremal Reissner-Nordstrom black holes. The tunneling amplitude for this instanton was found to be proportional to the difference in entropy between the initial and final black hole configurations. The probe limit of this instanton was studied in \cite{AdSFrag}, in which a single $AdS_2\times S_2$ space of charge $\Gamma_1+\Gamma_2$ with $\Gamma_1 \ggg \Gamma_2$ splits into an $AdS_2\times S_2$ space of charge $\Gamma_1$ with a probe charge $\Gamma_2$ in the boundary. This instanton was found to give the same tunneling amplitude as the Brill instanton.

In this work we consider the multi-centered black hole scaling solutions of the bosonic sector of $\mathcal{N}=2$ supergravity coupled to an arbitrary number of vector multiplets. We find an instanton analogous to the Brill instanton, that connects a single-centered scaling configuration of charge $\Gamma_\infty = \Gamma_1 + \Gamma_2 + \Gamma_3$ to three disconnected scaling throats of charges $\Gamma_1, \Gamma_2, \Gamma_3$, in the limit where the intersection products between the charges of the three throats satisfies $\langle \Gamma_1, \Gamma_2\rangle \lll \langle \Gamma_2, \Gamma_3\rangle, \langle \Gamma_3, \Gamma_1\rangle$. Thus, this instanton can be interpreted as describing the topological fluctuations of a single-centered scaling throat. We evaluate the Euclidean action for this instanton and find that it is proportional to the difference in entropy between the two black hole configurations.

This paper is structured as follows. In Section \ref{sec-MCSol} we describe the general multi-centered black hole solution to $\mathcal{N}=2$ supergravity, as well as the scaling solutions. In Section \ref{sec-Instanton} we find the instanton solution corresponding to tunneling between single and multi-centered scaling configurations. In Section \ref{sec-Action} we evaluate the Euclidean action for the instanton to find the tunneling amplitude. We conclude in Section \ref{sec-Conc}.

\section{The multi-centered black hole solution}\label{sec-MCSol}

The action for the bosonic part of four-dimensional $\mathcal{N}=2$ supergravity coupled to massless vector multiplets takes the form:
\begin{align}
S_{4D} &= \frac{1}{16\pi}\int_{M_4} d^4 x \sqrt{-g}\left (R - 2G_{A\bar{B}}dz^A \wedge \star d\bar{z}^{\bar{B}} - F^I \wedge G_I\right),
\end{align}
where the $z^A$ $(A=1,\dots,n)$ are the vector multiplet scalars, the $F^I$ $(I=0,1,\dots,n)$ are the vector field strengths, the $G_I$ are the dual magnetic field strengths, and $G_{A\bar{B}} = \partial_A\partial_{\bar{B}}\mathcal{K}$ is derived from the Kahler potential
\begin{align}
\mathcal{K} = -\ln (i\int_X \Omega_0\wedge\bar{\Omega}_0)
\end{align}
where $\Omega_0$ is the holomophic 3-form on the Calabi-Yau manifold $X$. For later calculations, it will also be convenient to define the normalized 3-form $\Omega = e^{\mathcal{K}/2}\Omega_0$.

The lattice of electric and magnetic charges $\Gamma$ is identified with $H^3(X,\mathbb{Z})$, the lattice of integral harmonic 3-forms on $X$. In the standard symplectic basis, a charge $\Gamma$ can be written as $\Gamma = (P^I, Q_I)$, with magnetic charges $P^I$ and electric charges $Q_I$. We can define a canonical, duality invariant, symplectic product $\langle, \rangle$ on the space of charges, which is given by:
\begin{align}\label{eq-IntProd}
\langle\Gamma, \tilde{\Gamma} \rangle = P^I\tilde{Q}_I - Q_I\tilde{P}^I
\end{align}
in the standard symplectic basis. For convenience of notation, we define $\Gamma_{ab} \equiv \langle \Gamma_a, \Gamma_b \rangle$. The central charge $Z_a$ of $\Gamma_a$ is given by:
\begin{align}
Z_a \equiv \langle \Gamma_a, \Omega \rangle.
\end{align}
In \cite{DenefSUGRA} it was shown that multicentered BPS black hole solutions for this theory exist, and their explicit form was found in \cite{DenefBates}. These solutions are regular, stationary, asymptotically flat BPS solutions with intrinsic angular momentum, describing bound states of separate extremal black holes with mutually nonlocal charges.

A general multicentered solution can be explicitly constructed given the entropy function $\Sigma$, which is a homogeneous function of degree two that is defined so that $\pi\Sigma(\Gamma) = S(\Gamma)$ is the entropy of a black hole of charge $\Gamma$. Given a multicentered configuration with $N$ black holes of charges $\Gamma_1, \dots, \Gamma_N$ at locations $\textbf{x}_1,\dots, \textbf{x}_N$, we define the harmonic function:
\begin{align}
H &= \sum_{a=1}^N \frac{\Gamma_a}{|\textbf{x} - \textbf{x}_a|} - 2\mathrm{Im}(e^{-i\alpha}\Omega)_{r=\infty}
\end{align}
The positions of the black holes are required to satisfy the integrability constraints:
\begin{align}
\sum_{a < b} \frac{\Gamma_{ab}}{|\textbf{x}_a - \textbf{x}_b|} = 2\mathrm{Im}(e^{-i\alpha}Z_a)_{r=\infty}\label{eq-Integ}
\end{align}
The fields of the black hole solution can be written explicitly in terms of $\Sigma(H)$. The metric has the form:
\begin{align}
ds^2 = -e^{2U}(dt + \omega_j dx^j)^2 + e^{-2U}dx^jdx^j
\end{align}
where
\begin{align}
e^{-2U} = \Sigma(H).
\end{align}
The 1-form $\omega = \omega_j dx^j$ is given by:
\begin{align}\label{eq-OmegaAdd}
\omega = \sum_{a < b} \omega_{ab}
\end{align}
where $\omega_{ab}$ is the value of the 1-form for a configuration of two charges $\Gamma_a, \Gamma_b$, given by:
\begin{align}\label{eq-OmegaAB}
\omega_{ab} = \frac{\Gamma_{ab}}{2l} \left (\frac{l^2 - r^2}{(l^4 + r^4 - 2l^2r^2\cos2\theta)^{1/2}} + 1 - \cos\theta_1 + \cos\theta_2 \right ) d\phi
\end{align}
where $2l$ is the distance between the two charges, and $r,\theta,\phi$ are spherical coordinates defined with respect to the axis between the two charge centers, with the origin halfway between the charges. The angles $\theta_1, \theta_2$ are the angles with the $z$-axis in a spherical coordinate system with origin at $\textbf{x}_1$ resp. $\textbf{x}_2$, and are related to the central spherical coordinates by $r\cos\theta - r_1\cos\theta_1 = l = r_2\cos\theta_2 - r\cos\theta$. Asymptotically for $r\to\infty$, we have $\omega \approx \frac{\Gamma_{ab}}{r}\sin^2\theta d\phi = \frac{\Gamma_{ab}}{r^3}(xdy - ydx)$.

The scalars $z^A$ are given by:
\begin{align}
z^A &= \frac{H^A-i\partial_{H_A}\Sigma}{H^0-i\partial_{H_0}\Sigma}
\end{align}
and the electromagnetic gauge fields are given by:
\begin{align}
\mathcal{A}^I &= \partial_{H_I}\log \Sigma(\mathrm{d}t + \omega) + \mathcal{A}_d^I, 
\end{align}
with
\begin{align}
\mathcal{A}_d = -\sum_a \Gamma_a \cos \theta_a \mathrm{d}\phi_a,
\end{align}
where spherical coordinates $(r_a,\theta_a,\phi_a)$ have been defined around each black hole center $\textbf{x}_a$.

As stated above, these multicentered solutions can be interpreted as bound states of separate extremal black holes with charges $\Gamma_1,\dots,\Gamma_N$. To see this, note that if we define $r_a \equiv |\textbf{x} - \textbf{x}_a|$, then in the near-horizon region $r_a \to 0$ of the black hole at $\textbf{x}_a$, we have
\begin{align}
\lim_{r_a \to 0} \Sigma(H) = \pi S(\Gamma_a)r_a^2.
\end{align}
Thus we see that if we ignore the $\omega_i$ cross-terms in the metric, then each of the black holes has the standard $AdS_2\times S_2$ near-horizon geometry of all four-dimensional extremal black holes.

These solutions have an intrinsic angular momentum $\vec{J}$, which can be extracted from the off-diagonal terms in the metric in the limit $r\to\infty$:
\begin{align}\label{eq-omegaInf}
\omega = 2\epsilon_{ijk}\frac{J^ix^jdx^k}{r^3} + \mathcal{O}(r^{-2})
\end{align}
In order to determine $\vec{J}$, we can use the asymptotic form of $\omega_{ab}$ for a two-centered solution:
\begin{align}
\omega_{ab} = \frac{\Gamma_{ab}}{r^3}(xdy - ydx) + \mathcal{O}(r^{-2})
\end{align}
Using the additivity of $\omega$ as shown in (\ref{eq-OmegaAdd}), we find that:
\begin{align}\label{eq-J}
\vec{J} &= \frac{1}{2}\sum_{a<b}\frac{\Gamma_{ab}(\vec{\textbf{x}}_a - \vec{\textbf{x}}_b)}{|\textbf{x}_a - \textbf{x}_b|}.
\end{align}

\subsection{The Scaling Solution}\label{sec-ScSol}

Scaling solutions are special instances of multicentered solutions where the black hole centers $\textbf{x}_a$ can approach arbitrarily close to each other. We can therefore write
\begin{align}
\textbf{x}_a = \lambda \textbf{w}_a + \lambda^2\textbf{s}_a + O(\lambda^3),
\end{align}
and let $\lambda \to 0$. The exact scaling solution is obtained in the limit $\lambda = 0$.

Naively, it might appear that as $\lambda \to 0$ the black holes collapse on top of each other. However, this is not the case. As $\lambda \to 0$, the black hole centers become encapsulated in a throat, which becomes deeper and deeper as $\lambda$ decreases. Due to the warp factor of the throat, the black holes remain at a finite physical distance from each other even while the coordinate distances $|\textbf{x}_a - \textbf{x}_b|$ become infinitely small. In the full scaling limit, this throat becomes an $AdS_2\times S_2$ space that can be considered as its own separate spacetime, and the solution gains a scaling symmetry. From outside the $AdS_2\times S_2$ throat, the configuration looks like the throat of a single black hole with total charge $\Gamma_\infty = \Gamma_1 + \dots + \Gamma_N$.

As $\Sigma$ is homogeneous of degree 2, in the full scaling limit $\lambda = 0$, the metric becomes purely spatial, and $g_{tt} \to 0$. To avoid this, we rescale the time coordinate $t \to t/\lambda$. We can then write the solution in terms of the scaling coordinates $\textbf{w}$ and the new $t$-coordinate:
\begin{align}
ds^2 = -e^{2U}(dt + \omega_j dw^j)^2 + e^{-2U}dw^jdw^j,
\end{align}
where now $U$ and $\omega_j$ are functions of $\textbf{w}$. Upon studying the equations of motions for the metric, the scalar fields, and the electromagnetic fields, it turns out that all of the equations are exactly the same as in the non-scaling case except with the variable $\textbf{w}$ substituted for $\textbf{x}$, and also with the constant in the harmonic function $H(w)$ set to 0. Thus the positions of the black holes now satisfy the integrability constraints:
\begin{align}\label{eq-Integ2}
\sum_{a < b} \frac{\Gamma_{ab}}{|\textbf{w}_a - \textbf{w}_b|} = 0.
\end{align}
For convenience of notation, we will now switch from using the $\textbf{w}$ coordinate to just using the $\textbf{x}$ coordinate, with the understanding that the constant in the harmonic function is set to 0.

\subsection{The 3-centered Scaling Solution}\label{sec-Approx}

In this work we consider scaling solutions with three black holes. Due to the integrability constraints (\ref{eq-Integ2}), the positions of the black holes are given by:
\begin{align}\label{eq-TriSides}
|\textbf{x}_a - \textbf{x}_b| = |\Gamma_{ab}|\rho
\end{align}
for some positive parameter $\rho$. Thus the three $\Gamma_{ab}$ correspond to the lengths of the three sides of a triangle. We can label the charges so that:
\begin{align}
\Gamma_{12} &> 0\\
\Gamma_{31} &> 0\\
\Gamma_{23} &> 0.
\end{align}
As the $\Gamma_{ab}$ are proportional to the distances between the black holes, we see from the formula (\ref{eq-J}) that the intrinsic angular momentum of this spacetime is zero, which means that $\omega = \mathcal{O}(r^{-2})$ for any three-centered scaling solution, so that $\omega_r \sim \frac{1}{r^3}$ and $\omega_\theta, \omega_\phi \sim \frac{1}{r^2}$ as $r \to \infty$.

In this work, we consider the limit in which the charges $\Gamma_1, \Gamma_2,\Gamma_3$ satisfy:
\begin{align}\label{eq-3Approx}
\frac{\Gamma_{12}}{\Gamma_{31}}, \frac{\Gamma_{12}}{\Gamma_{23}}\to 0
\end{align}
In this limit, the interaction between the two black holes of charge $\Gamma_1, \Gamma_2$, characterized by the quantity $\Gamma_{12}$, is small compared to the interaction between each of $\Gamma_1$, $\Gamma_2$ and the third black hole of charge $\Gamma_3$, characterized by the quantities $\Gamma_{23}, \Gamma_{31}$. Note, however, that by the definition of the intersection product given in (\ref{eq-IntProd}), this does not necessarily mean that the charges $\Gamma_1$, $\Gamma_2$ are small compared to $\Gamma_3$.

This limit ensures that $\omega \to 0$ as $\textbf{x} \to \textbf{x}_a$, so that the geometry at each of the black holes $\textbf{x}_a$ is that of an $AdS_2 \times S_2$ throat of charge $\Gamma_a$. In this limit, we also find that $\omega \to 0$ for $|\textbf{x}| \ggg \Gamma_{31}\rho$ (for details, see Appendix \ref{sec-App}.) We will see later that this is necessary in order to satisfy the conditions for a valid instanton solution: namely, that the metric is real in the regions corresponding to asymptotically early and late Euclidean times. However, this limit will \textit{not} be necessary in order to actually evaluate the Euclidean action for the instanton.

\section{The instanton solution}\label{sec-Instanton}


\subsection{The Brill instanton and $AdS_2$ fragmentation}\label{sec-Brill}

The fragmentation process that we are considering is similar to the fragmentation of the $AdS_2 \times S_2$ throats of extremal Reissner-Nordstrom (RN) black holes that was studied in \cite{Brill} and \cite{AdSFrag}. It will be helpful to first study this simpler example before considering the full $\mathcal{N}=2$ SUGRA case.

The analog of our multicentered scaling solution is the class of \textit{conformastatic} solutions to the Einstein-Maxwell equations. The Lorentzian conformastatic metric and Maxwell field have the form:
\begin{align}
ds^2 &= -H^{-2}dt^2 + H^2dx^jdx^j\\
\star F &= dt \wedge dH,
\end{align}
where $H$ is a harmonic function that satisfies
\begin{align}
\nabla^2 H = 0,
\end{align}
where $\nabla^2$ is the Laplacian on flat $\mathbb{R}^3$. This solution describes a Bertotti-Robinson (BR) type universe\cite{BertRob} containing a number of extremal Reissner-Nordstrom (ERN) black holes. If the black holes are located at coordinates $\textbf{x}_a$, then the function $H$ has the general form:
\begin{align}
H = \sum_{a=1}^N \frac{\Gamma_a}{|\textbf{x} - \textbf{x}_a|},
\end{align}
where $\Gamma_a$ are the charges of the black holes. The spacetime extends over the range $t \in \{-T, T\}$, $r > R$, and $r_a > R_a$, where $r_a \equiv |\textbf{x} - \textbf{x}_a|$. In the end we take the limits $T, R \to \infty$ and $R_a\to 0$.

Brill found an instanton that describes the fragmentation of one $AdS_2\times S_2$ universe into several by analytically continuing this solution to Euclidean time. He identified two asymptotic regions corresponding to the initial and final states: in the limit $r = |\textbf{x}|\to\infty$, the metric approaches an $AdS_2\times S_2$ throat of a single black hole with charge $\Gamma_\infty = \Gamma_1 + \dots + \Gamma_N$. In the limit $r_a = |\textbf{x} - \textbf{x}_a|\to 0$, the metric approaches an $AdS_2\times S_2$ throat of a black hole with charge $\Gamma_a$.

Thus, we can define Euclidean time by taking the level surfaces of the harmonic function $H$ to be a foliation of the spacetime. For early times, corresponding to $H\to 0$ (and thus $r\to\infty$), we have a single $AdS_2\times S_2$ throat of charge $\Gamma_\infty$. For late times, corresponding to $H\to\infty$ (and thus $r_a \to 0$), we have $N$ separate $AdS_2\times S_2$ throats of charges $\Gamma_1,\dots,\Gamma_N$. As the initial and final geometries are only reached asymptotically in Euclidean time, this is not a ``bounce'' instanton corresponding to a genuine decay out of the initial state. Rather, it is analogous to the well-known instanton solution for the symmetric double well, which indicates mixing between two degenerate minima. Upon evaluating the Euclidean action, we find that the tunneling amplitude for the single $AdS_2\times S_2$ universe to fragment into multiple $AdS_2\times S_2$ universes is given by:
\begin{align}\label{eq-BrillAmp}
\frac{\pi}{2} \left ( \Gamma_\infty^2 - \sum_{a=1}^N \Gamma_a^2\right )
\end{align}
The Brill instanton does not connect a single-centered conformastatic solution to a multi-centered conformastatic solution: rather, it connects a single-centered solution to a geometry containing $N$ disconnected $AdS_2\times S_2$ centers, which agrees with the multi-centered conformastatic configuration deep inside the black hole throats. Note that the tunneling amplitude for the instanton is proportional to the entropy difference between these initial and final black hole configurations. Brill conjectured that an analogous instanton with the same (or similar) amplitude would exist that would connect the full single and multi-centered conformastatic solutions, and thus describe the splitting of a single ERN black hole into multiple black holes; however, such an instanton has not yet been found.

The probe limit of the two-centered Brill instanton was studied in \cite{AdSFrag}, where a probe charge $\Gamma_2$ was considered in the $AdS_2\times S_2$ throat of a large background ERN black hole of charge $\Gamma_1 \ggg \Gamma_2$. An instanton was found that connected a single $AdS_2\times S_2$ space of charge $\Gamma_1+\Gamma_2$ to an $AdS_2\times S_2$ space of charge $\Gamma_1$ with a probe charge $\Gamma_2$ in the boundary of the $AdS_2$ space. The tunneling amplitude for the instanton was found to be proportional to the entropy difference between the initial and final configurations. The instanton is the probe limit of the Brill instanton.

\subsection{The $\mathcal{N}=2$ SUGRA instanton}

The $\mathcal{N}=2$ SUGRA instanton can be obtained by analytically continuing the existing stationary solution to imaginary time. Wick-rotation gives the following Euclidean metric:
\begin{align}
ds^2 = e^{2U}(dt - i\omega_j dx^j)^2 + e^{-2U}dx^jdx^j
\end{align}
As in the RN case, our spacetime extends over the range $t \in \{-T, T\}$, $r > R$, and $r_a > R_a$, where $r_a \equiv |\textbf{x} - \textbf{x}_a|$. In the end we take the limits $T, R \to \infty$ and $R_a\to 0$.

Recall that we are considering the particular three-centered configuration described in Sec. \ref{sec-Approx}, in the limit $\frac{\Gamma_{12}}{\Gamma_{31}}, \frac{\Gamma_{12}}{\Gamma_{23}} \to 0$. For this solution, we find that $\omega\to 0$ as $r_a\to 0$ and as $r\to\infty$ (for explicit calculations, see Appendix \ref{sec-App1}.) 

We define Euclidean time by taking the level surfaces of the entropy function $\Sigma(H)$ to be a foliation of the spacetime. As shown above, we find that $\omega \to 0$ for $r\to\infty$ and $r_a\to 0$. Thus, for early times, corresponding to $\Sigma\to 0$ (and $r\to\infty$), we have a single $AdS_2\times S_2$ throat of charge $\Gamma_\infty = \Gamma_1 + \Gamma_2 + \Gamma_2$. For late times, corresponding to $\Sigma\to\infty$ (and $r_a\to 0$), we have three separate $AdS_2\times S_2$ throats of charges $\Gamma_1,\Gamma_2,\Gamma_3$. As $\omega \to 0$ for $r \to \infty$ and $r_a \to 0$, the metric is real in the asymptotic regions. Thus, as long as the Euclidean action is real when evaluated on this solution, this is a valid gravitational instanton.

As with the Brill instanton, our instanton connects a single-centered scaling solution to a geometry containing three disconnected scaling throats, which agrees with the three-centered scaling configuration deep inside the black hole throats. Thus, using the analogy with the Brill instanton, it is reasonable to say that this instanton describes topological fluctuations of the single-centered scaling throat, and the Euclidean action for this instanton gives the tunneling amplitude for these fluctuations. In the probe limit, our instanton corresponds to the escape of two probe particles of charges $\Gamma_1, \Gamma_2$ from the $AdS_2\times S_2$ throat of a single large black hole of charge $\Gamma_3$ (though as mentioned in Section \ref{sec-Approx}, this does not necessarily mean that the charges $\Gamma_1$, $\Gamma_2$ are small compared to $\Gamma_3$, but that the interaction between first two charges is small compared to their interactions with the third charge.)

\section{Evaluating the Euclidean action}\label{sec-Action}

We can now evaluate the Euclidean action. The form of the action for a general stationary solution is\cite{DenefSUGRA}:
\begin{align}
S_{4D} = -\frac{1}{16\pi} \int dt \int_{\mathbb{R}^3} \{ 2\spd{U}\wedge\sph\spd U - \frac{1}{2}e^{4U}\spd \omega\wedge\sph\spd\omega + 2G_{A\bar{B}}\spd z^A \wedge \sph\spd \bar{z}^{\bar{B}} + (\mathcal{F},\mathcal{F})  \}
\end{align}
where $\spd$ and $\sph$ are the exterior derivative and the three-dimensional Hodge dual with respect to \textit{flat} Euclidean space respectively, $\mathcal{F} = \mathcal{F}_{ij}$ is the spatial part of the electromagnetic field, and the scalar product $(,)$ of spatial 2-forms $\mathcal{F}$ and $\mathcal{F}'$ is defined as:
\begin{align}
(\mathcal{F},\mathcal{F}') \equiv \frac{e^{2U}}{1-\tilde{\omega}^2}\int_X \mathcal{F}\wedge \left [\sph\hat{\mathcal{F}'} - \sph(\tilde{\omega}\wedge\hat{\mathcal{F}'})\tilde{\omega} + \sph(\tilde{\omega}\wedge\sph\mathcal{F}') \right ]
\end{align}
with $\tilde{\omega} \equiv e^{2U}\omega$. In \cite{DenefSUGRA} it was shown that the integrand could be rewritten as:
\begin{align}
\mathcal{L} &= (\mathcal{G},\mathcal{G}) - 4(\mathcal{Q} + \spd\alpha + \frac{1}{2}e^{2U}\sph\spd\omega)\wedge \mathrm{Im} \langle\mathcal{G},e^Ue^{-i\alpha}\Omega \rangle\nonumber\\
&\quad+\spd [ 2\tilde{\omega}\wedge(\mathcal{Q}+\spd\alpha) + 4\mathrm{Re}\langle \mathcal{F},e^Ue^{-i\alpha}\Omega \rangle]
\end{align}
where $\alpha$ is an arbitrary real function on $\mathbb{R}^3$ and $\mathcal{Q}$ is the spatial part of the chiral connection:
\begin{align}
\mathcal{Q} = \mathrm{Im}(\partial_A\mathcal{K}\spd z^A).
\end{align}
The 2-form $\mathcal{G}$ is defined as:
\begin{align}
\mathcal{G} \equiv \mathcal{F} - 2\mathrm{Im}\sph \textbf{D}(e^{-U}e^{-i\alpha}\Omega) + 2\mathrm{Re}\textbf{D}(e^Ue^{-i\alpha}\Omega\omega).
\end{align}
and
\begin{align}
\textbf{D} \equiv \textbf{d} + i(\mathcal{Q} + \textbf{d}\alpha + \frac{1}{2}e^{2U}\sph \textbf{d}\omega).
\end{align}
For the scaling solution, we have\cite{DenefSUGRA}
\begin{align}
\mathcal{Q} + \textbf{d}\alpha &= \frac{1}{2}e^{2U}\langle\spd H, H\rangle\label{eq-UsefulId1}\\
\spd U &= e^U\mathrm{Re}(e^{-i\alpha}\zeta)\label{eq-UsefulId2}\\
\sph \spd \omega &= \langle \spd H, H\rangle\label{eq-UsefulId3}\\
\mathcal{F} &= 2\sph\spd \mathrm{Im} (e^{-U}e^{-i\alpha}\Omega) - 2\spd\mathrm{Re}(e^Ue^{-i\alpha}\Omega\omega)\label{eq-UsefulId4}
\end{align}
where
\begin{align}\label{eq-zeta}
\zeta \equiv \langle\spd H, \Omega \rangle = \sum_{i=1}^N Z(\Gamma_i)\spd\left (\frac{1}{|\textbf{x}-\textbf{x}_i|}\right ).
\end{align}
In addition to the boundary terms
\begin{align}
&\spd [2\tilde{\omega}\wedge (\mathcal{Q} + \spd\alpha)]\\
&\spd [4 \mathrm{Re} \langle \mathcal{F}, e^U e^{-i\alpha}\Omega\rangle]
\end{align}
given above, the action also has the following boundary term:
\begin{align}
&2 \nabla U,
\end{align}
where the Laplacian is taken with respect to three dimensional \textit{flat} space. The term $\sim\nabla U$ was omitted in \cite{DenefSUGRA} because it did not contribute in asymptotically flat space, but since the scaling solution is asymptotically $AdS_2\times S_2$ it must be included here. The boundary terms in the action also include the Gibbons-Hawking term, required for a valid action principle for the metric. 

When evaluated on a solution of the equations of motion, the bulk term is zero. Thus the only contribution to the Euclidean action comes from the boundary terms.

\subsection{The Gibbons-Hawking term}

The Gibbons-Hawking term for the Euclidean solution is given by\cite{HawkingIsrael}:
\begin{align}
-\frac{1}{8\pi}\int d^3 x \,\,\sqrt{h}\, K  + C[h_{ij}]
\end{align}
where $h_{ij}$ is the induced metric on the boundary, $K$ is the trace of the second fundamental form of the boundary, and $C[h_{ij}]$ is a term that depends solely on the induced metric at the boundary. Recall that our spacetime extends over the range $t \in \{-T,T\}$, $r > R$, and $r_a > R_a$, where $r_a \equiv |\textbf{x} - \textbf{x}_a|$. We have three boundaries to consider:
\begin{enumerate}
\item The top and bottom surfaces: $t = T$, $r<R$, $r_a > R_a$, and $t = -T$, $r<R$, $r_a > R_a$.

\item The ``mantle'' surfaces: $-T < t < T$, $r=R$ and $-T < t < T$, $r_a = R_a$.

\item The ``edges'': $t = \pm T$, $r=R$, $r_a = R_a$.
\end{enumerate}
On the top surface we have:
\begin{align}
K &= \frac{i}{2\Sigma^{3/2}(\Sigma^2 - \omega^2)}\{ -\Sigma^2 (\omega_i\partial_i\Sigma) - \omega^2(\omega_i\partial_i\Sigma)\nonumber\\
&\qquad+ 2\Sigma^2(\partial_i\omega_i) - 2\Sigma(\omega_x^2(\partial_z\omega_z+\partial_y\omega_y) +\omega_y^2(\partial_x\omega_x+\partial_z\omega_z)+\omega_z^2(\partial_x\omega_x+\partial_y\omega_y)\nonumber\\
&\qquad- \omega_x\omega_y(\partial_y\omega_x+\partial_x\omega_y) -\omega_x\omega_z(\partial_z\omega_x+\partial_x\omega_z)-\omega_y\omega_z(\partial_y\omega_z+\partial_z\omega_y)) \}\\
&= \frac{i}{2\Sigma^{3/2}(\Sigma^2 - \omega^2)}\{ -\Sigma^2 (\omega_i\partial_i\Sigma) - \omega^2(\omega_i\partial_i\Sigma)\nonumber\\
&\qquad- 2\Sigma(\omega_x^2(\partial_z\omega_z+\partial_y\omega_y) +\omega_y^2(\partial_x\omega_x+\partial_z\omega_z)+\omega_z^2(\partial_x\omega_x+\partial_y\omega_y)\nonumber\\
&\qquad- \omega_x\omega_y(\partial_y\omega_x+\partial_x\omega_y) -\omega_x\omega_z(\partial_z\omega_x+\partial_x\omega_z)-\omega_y\omega_z(\partial_y\omega_z+\partial_z\omega_y)) \}
\end{align}
using $\partial_i\omega_i = 0$. And
\begin{align}
\sqrt{h} = r^2\sin\theta\Sigma^{3/2}
\end{align}

On the ``mantle'' surfaces we have:
\begin{align}
K = \frac{4\Sigma + r\partial_r\Sigma}{2r\Sigma^{3/2}}
\end{align}
and
\begin{align}
\sqrt{h} = r^2\sin\theta\sqrt{\Sigma}
\end{align}
In the limit $R\to\infty$, on the ``outer mantle'' surface we have:
\begin{align}
K \to \frac{1}{\Sigma(\Gamma_\infty)}
\end{align}
and
\begin{align}
\sqrt{h} \to r\sin\theta\sqrt{\Sigma(\Gamma_\infty)}
\end{align}
where $\Gamma_\infty = \Gamma_1 + \Gamma_2 + \Gamma_3$. In the limit $R_a\to 0$, on the ``inner mantle'' surface we have:
\begin{align}
K \to -\frac{1}{\Sigma(\Gamma_a)}
\end{align}
and
\begin{align}
\sqrt{h} \to r\sin\theta\sqrt{\Sigma(\Gamma_a)}
\end{align}
The difference in sign arises from the normals on the inner and outer mantle surfaces pointing in different directions.

We take the limits $R \to \infty$, $R_a \to 0$ before taking $T\to \infty$. We can then take $C[h_{ij}]$ to be such that it cancels the contribution from the outer mantle surface, and the top and bottom surfaces. This is possible because for both the outer mantle surface, and the top and bottom surfaces, $\sqrt{h}K$ only depends on the induced metric $h_{ij}$ at the boundary. Thus the Gibbons-Hawking term will be zero in the case of a scaling solution with just one throat, where $\Gamma_\infty = \Gamma_1$ and all other $\Gamma_a = 0$. The contributions from the inner mantle surfaces is zero since $R_a\to 0$.

Finally, the contribution from the ``edges'' can be calculated using the results in \cite{Hayward}: the contribution from the $a$th pair of edges is $\pi A_a$, where $A_a$ is the area of the edge $R_a \to 0$, which is $4\pi \Sigma(\Gamma_a)$. The contribution to the Gibbons-Hawking term from an edge formed by two boundaries with spacelike normals $n_0, n_1$ (which is the case here, in Euclidean spacetime) is given by:
\begin{align}
\int_{edge} -i\eta \sigma^{1/2}d^2x = -i\eta A_a
\end{align}
where the factor of $-i$ comes from Wick-rotation, and we have defined
\begin{align}
\eta\equiv \mathrm{arccosh}(-n_0\cdot n_1)
\end{align}
In this case we have:
\begin{align}
\mathrm{arccosh}(-n_0\cdot n_1) &\sim \mathrm{arccosh}\left(\frac{1}{\Sigma}i\omega_r\right) \to \frac{i\pi}{2}
\end{align}
as $\omega_r/\Sigma \to 0$ at the edges (this is the case even when we do not take the limit $\frac{\Gamma_{12}}{\Gamma_{31}}, \frac{\Gamma_{12}}{\Gamma_{23}} \to 0$.) This is obvious for the ``inner'' edges $r_a=R_a$, as $\omega_r/\Sigma \sim R_a^2 \to 0$ as $r_a \to 0$. At the ``outer'' edges $r=R$, we have:
\begin{align}
\frac{1}{\Sigma}i\omega_r &\sim iR^2\frac{1}{R^3} \to 0
\end{align}
as $R \to \infty$.

Since $n_0\cdot n_1 \to 0$ at all the edges, the directions of the normals (i.e. whether they are inward or outward-pointing) does not matter, since we can take the branch of arccosh such that
\begin{align}
\mathrm{arccosh} 0 = \frac{i\pi}{2}
\end{align}
whether 0 is approached from above or below. Thus each edge will give a contribution of the same sign.

As with the contribution from the mantle surfaces, we want to normalize the Euclidean action so that it is zero when evaluated on a single-centered scaling solution of total charge $\Gamma_\infty = \Gamma_1$. Without normalizing, the edge terms for such a solution add up to:
\begin{align}
\frac{1}{8\pi}(\pi A_1 + \pi A_\infty) = \frac{1}{4} A_\infty,
\end{align}
where the first term comes from the pair of edges $r_1=R_1\to 0$, $t=\pm T$, and the second term comes from the pair of edges $r=R\to\infty$, $t=\pm T$. So we must subtract $\frac{1}{4} A_\infty$ to obtain the correct normalization. This means that the total contribution to the Euclidean action from the edge terms is:
\begin{align}
-\left(\frac{1}{8\pi}\sum_{a=1}^N \pi A_a + \pi A_\infty - \frac{1}{4}A_\infty \right)&= -\frac{1}{8}\sum_{a=1}^N A_a +\frac{1}{8} A_\infty\\
&= \frac{\pi}{2} \left ( \Sigma(\Gamma_\infty) -\sum_{a=1}^N \Sigma(\Gamma_a)\right)
\end{align}
This is proportional to the difference in entropy between the initial and final configurations.

\subsection{The remaining boundary terms}

All of the remaining boundary terms may be evaluated solely at the \textit{spatial} boundary, i.e. the ``mantle'' surfaces, as they involve the spatial exterior derivative $\spd$. So we use:
\begin{align}
\int_{\mathbb{R}^3} \spd A = \int_{\partial\mathbb{R}^3} A
\end{align}
where $\partial\mathbb{R}^3$ are the surfaces $-T < t < T$, $r=R$ and $-T < t < T$, $r_a = R_a$.

First consider the term:
\begin{align}
\spd [2\tilde{\omega}\wedge (\mathcal{Q} + \spd\alpha)]
\end{align}
This boundary term does not contribute to the final Euclidean action (once again, this is the case even when we do not take the limit $\frac{\Gamma_{12}}{\Gamma_{31}}, \frac{\Gamma_{12}}{\Gamma_{23}} \to 0$.) This is easy to see for the boundaries at the black holes, $r_a=R_a$, since $\tilde{\omega}=e^{2U}\omega$ and $e^{2U}\sim r_a^2$ at the $a$th black hole which goes to zero as $r_a \to 0$. At the outer boundary, $r=R$, using (\ref{eq-UsefulId1}) to substitute for $(\mathcal{Q} + \spd\alpha)$, we have:
\begin{align}
[2\tilde{\omega}\wedge (\mathcal{Q} + \spd\alpha)]_{\theta\phi} &\sim e^{2U}\omega_{\theta} \left (e^{2U}\langle \spd H, H\rangle \right )_{\phi} \quad\mathrm{or}\quad e^{2U}\omega_{\phi} \left (e^{2U}\langle \spd H, H\rangle \right )_{\theta}\\
&\sim R^2 \frac{1}{R^2} R^2\frac{1}{R^3} \to 0
\end{align}
as $R\to \infty$.

Next consider the term:
\begin{align}
&2 \nabla U
\end{align}
The contribution of this boundary term is given by:
\begin{align}
\int_{\mathbb{R}^3} 2 \nabla U &= \int_{\partial\mathbb{R}^3} 2 \spd U \cdot d\vec{a}
\end{align}
As $e^{-2U} = \Sigma$, we can evaluate this term using:
\begin{align}
\partial_r U = -\frac{1}{2}\frac{\partial_r \Sigma}{\Sigma}.
\end{align}
On the outer mantle surface, the boundary term is:
\begin{align}
\int_{\partial\mathbb{R}^3} 2 \spd U \cdot \spd a &= -\int_{r=R}r^2\sin\theta d\theta d\phi\frac{\partial_r \Sigma}{\Sigma} \to 8\pi R \Sigma(\Gamma_\infty)
\end{align}
On the inner mantle surface, the boundary term is:
\begin{align}
\int_{\partial\mathbb{R}^3} 2 \spd U \cdot \spd a &= \int_{r_a=R_a}r^2\sin\theta d\theta d\phi\frac{\partial_r \Sigma}{\Sigma} \to -8\pi R_a \Sigma(\Gamma_a)
\end{align}
with the sign difference being due to the normal being either inward or outward-pointing.

Finally, in order to evaluate the boundary term:
\begin{align}
&\spd [4 \mathrm{Re} \langle \mathcal{F}, e^U e^{-i\alpha}\Omega\rangle],
\end{align}
we use (\ref{eq-UsefulId4}) to write:
\begin{align}
\mathcal{F} &= 2\sph\spd \mathrm{Im} (e^{-U}e^{-i\alpha}\Omega) - 2\spd\mathrm{Re}(e^Ue^{-i\alpha}\Omega\omega)\\
&= -\sph\spd H - 2\spd\mathrm{Re}(e^Ue^{-i\alpha}\Omega\omega)
\end{align}
So we can write:
\begin{align}
\mathrm{Re} \langle \mathcal{F}, e^U e^{-i\alpha}\Omega\rangle &= \langle \mathcal{F}, \mathrm{Re} (e^U e^{-i\alpha}\Omega)\rangle\\
&= \langle -\sph\spd H - 2\spd\mathrm{Re}(e^Ue^{-i\alpha}\Omega\omega), \mathrm{Re} (e^U e^{-i\alpha}\Omega)\rangle\\
&= \langle-\sph\spd H,\mathrm{Re} (e^U e^{-i\alpha}\Omega) \rangle - 2\langle \spd\mathrm{Re}(e^Ue^{-i\alpha}\Omega\omega),\mathrm{Re} (e^U e^{-i\alpha}\Omega)\rangle\\
&= -\sph\langle \spd H, \mathrm{Re} (e^U e^{-i\alpha}\Omega) \rangle - 2\langle \spd\mathrm{Re}(e^Ue^{-i\alpha}\Omega\omega),\mathrm{Re} (e^U e^{-i\alpha}\Omega)\rangle\\
&= -\sph \mathrm{Re}(e^Ue^{-i\alpha}\zeta) - 2\langle \spd\mathrm{Re}(e^Ue^{-i\alpha}\Omega\omega),\mathrm{Re} (e^U e^{-i\alpha}\Omega)\rangle.
\end{align}
where $\zeta$ is defined in (\ref{eq-zeta}). Using (\ref{eq-UsefulId2}) to substitute for the first term, we can write:
\begin{align}\label{eq-EMBound}
\mathrm{Re} \langle \mathcal{F}, e^U e^{-i\alpha}\Omega\rangle &= -\sph\spd U - 2\langle \spd\mathrm{Re}(e^Ue^{-i\alpha}\Omega\omega),\mathrm{Re} (e^U e^{-i\alpha}\Omega)\rangle
\end{align}
We can then evaluate
\begin{align}
\int_{\mathbb{R}^3} \spd [4 \mathrm{Re} \langle \mathcal{F}, e^U e^{-i\alpha}\Omega\rangle] = \int_{\partial\mathbb{R}^3} 4 \mathrm{Re} \langle \mathcal{F}, e^U e^{-i\alpha}\Omega\rangle
\end{align}
The second term in (\ref{eq-EMBound}) will go to zero at both mantle surfaces (once again, this is the case even when we do not take the limit $\frac{\Gamma_{12}}{\Gamma_{31}}, \frac{\Gamma_{12}}{\Gamma_{23}} \to 0$.) This is clearly the case at the inner mantle surface $r_a = R_a$ as $e^{2U} \to r^2_a \to 0$ and $\omega$ remains finite. At the outer mantle surface $r = R$, using (\ref{eq-UsefulId2}) and (\ref{eq-UsefulId3}) to evaluate $\spd U$ and $\spd \omega$, we see that as $R\to\infty$, we have
\begin{align}
\spd U &\to \frac{1}{R}\\
\spd \omega &\to \frac{1}{R^3}.
\end{align}
Thus the order of the 2nd term will be at most:
\begin{align}
\langle \spd\mathrm{Re}(e^Ue^{-i\alpha}\Omega\omega),\mathrm{Re} (e^U e^{-i\alpha}\Omega)\rangle &\sim \frac{1}{R} \to 0
\end{align}
as $R\to \infty$. So we get:
\begin{align}
\int_{\partial\mathbb{R}^3} 4 \mathrm{Re} \langle \mathcal{F}, e^U e^{-i\alpha}\Omega\rangle = -\int_{\partial\mathbb{R}^3} 4\sph\spd U
\end{align}
On the outer mantle surface, this is:
\begin{align}
\int_{\partial\mathbb{R}^3} -4\sph\spd U &= 4\int_{r=R}r^2\sin\theta d\theta d\phi\frac{1}{2}\frac{\partial_r \Sigma}{\Sigma} \to -8\pi R \Sigma(\Gamma_\infty)
\end{align}
On the inner mantle surface, this is:
\begin{align}
\int_{\partial\mathbb{R}^3} 4\sph\spd U &= -4\int_{r_a=R_a}r^2\sin\theta d\theta d\phi\frac{1}{2}\frac{\partial_r \Sigma}{\Sigma} \to 8\pi R_a \Sigma(\Gamma_a)
\end{align}
with the sign difference being due to the normal being either inward or outward-pointing. These terms cancel exactly with the boundary contribution from the $2\nabla U$ term, and thus these two terms do not contribute to the Euclidean action.

\subsection{The final value of the Euclidean action}

If the $C[h_{ij}]$ term in the Gibbons-Hawking term is chosen carefully, the only non-zero contribution to the Euclidean action comes from the ``edges'', and is equal to the difference in entropy between the initial and final configurations. Note that the probe limit $\frac{\Gamma_{12}}{\Gamma_{23}}, \frac{\Gamma_{12}}{\Gamma_{31}} \to 0$ was never required at any point in order to evaluate the Euclidean action. It was merely required in order to ensure that the Euclidean metric was real in the asymptotic regions.

\section{Conclusion}\label{sec-Conc}

We have found an instanton solution of $\mathcal{N}=2$ SUGRA that can be interpreted as a tunneling process from a single-centered black hole scaling throat of charge $\Gamma_\infty$ to three disconnected throats of charges $\Gamma_1,\Gamma_2,\Gamma_2$ such that $\Gamma_\infty = \Gamma_1 + \Gamma_2 + \Gamma_N$, in the limit $\frac{\Gamma_{12}}{\Gamma_{23}}, \frac{\Gamma_{12}}{\Gamma_{31}} \to 0$. The amplitude for the tunneling process is given by the difference in entropy between the initial and final configurations.

\section{Acknowledgements}

This work was funded in part by a Research Assistantship from Harvard's Center for the Fundamental Laws of Nature.

\begin{appendix}\label{sec-App}

\section{The behavior of the metric components $\omega_i$ in the asymptotic regions}\label{sec-App1}

Here we give the explicit form of the 1-form $\omega$ in the asymptotic regions $r\to\infty$ and $r_a\to 0$. We first consider $r\to\infty$. As mentioned in Section \ref{sec-Approx}, we have $\omega = \mathcal{O}(r^{-2})$ for any three-centered scaling solution. In the particular limit we are taking, with
\begin{align}
\frac{\Gamma_{12}}{\Gamma_{31}}, \frac{\Gamma_{12}}{\Gamma_{23}} \to 0,
\end{align}
we can evaluate
\begin{align}
\omega = \omega_{12} + \omega_{13} + \omega_{23}
\end{align}
using the formula (\ref{eq-OmegaAB}) for $\omega_{ab}$. If we set up Cartesian coordinates so that $\textbf{x}_3$ is at the origin, $\textbf{x}_1$ lies on the $z$-axis, and $\textbf{x}_2$ in the $y-z$ plane, then in the limit $\frac{\Gamma_{12}}{\Gamma_{31}}, \frac{\Gamma_{12}}{\Gamma_{23}} \to 0$, we can explicitly solve for the $\omega_{ab}$ to find:
\begin{align}
\omega_{12} &= \frac{-2\Gamma_{12}}{r_1r_2(2l_{12} + r_1 + r_2)}\left ((z - 2l_{13}\cos\beta)dx - xdz \right )\\
\omega_{13} &= \frac{2\Gamma_{31}}{r_1r_3(2l_{13} + r_1 + r_3)}\left ((y\cos\beta + z\sin\beta)dx - x(\cos\beta dy + \sin\beta dz) \right )\\
\omega_{23} &= \frac{-2\Gamma_{23}}{r_2r_3(2l_{23} + r_2 + r_3)}\left ((y\cos\beta - z\sin\beta)dx - x(\cos\beta dy - \sin\beta dz) \right )
\end{align}
where $r_a$ is the coordinate distance to $\Gamma_a$, given by $|\textbf{x} - \textbf{x}_a|$, the length $2l_{ab} = |\textbf{x}_a - \textbf{x}_b|$, and the angle $\beta$ satisfies:
\begin{align}
\sin\beta \sim \frac{\Gamma_{12}}{2\Gamma_{31}}
\end{align}
Explicit evaluation shows that for $r \ggg \Gamma_{31}\rho$, where $\rho$ is a finite positive parameter that gives the characteristic length scale of the scaling solution as defined in (\ref{eq-TriSides}), we have $\omega \sim \frac{\Gamma_{12}}{\Gamma_{31}} \to 0$.

We now consider the asymptotic regions $r_a \to 0$. From the form of $\omega_{ab}$ for two centers $\Gamma_a, \Gamma_b$, and the additivity of $\omega$, we see that at the black hole center $\textbf{x}_3$, the only non-zero contribution to $\omega$ is from  $\omega_{12}$. Similarly, at the center $\textbf{x}_1$, the only non-zero contribution to $\omega$ is from $\omega_{23}$, and at $\textbf{x}_2$, the only non-zero contribution to $\omega$ is from $\omega_{13}$.

At $\Gamma_3$, since we are considering the limit $\frac{\Gamma_{12}}{\Gamma_{31}}, \frac{\Gamma_{12}}{\Gamma_{23}} \to 0$, and the distances between the black holes are proportional to the $\Gamma_{ab}$ as given by (\ref{eq-TriSides}), we can use the long-distance approximation for $\omega_{12}$ to write
\begin{align}
\omega_{12} \sim \frac{\Gamma_{12}}{\Gamma_{31}\rho}\sin^2\theta d\phi,
\end{align}
where $\theta,\phi$ are defined with respect to the axis connecting the black hole centers $\textbf{x}_1$ and $\textbf{x}_2$. Thus $\omega \to 0$ as $r_3 \to 0$.

At $\Gamma_2$, we can explicitly evaluate $\omega_{13}$ using (\ref{eq-OmegaAB}) to find:
\begin{align}
\omega_{13} = \frac{\Gamma_{13}}{\Gamma_{31}\rho}\left (\frac{l^2 - r^2}{(l^4 + r^4 - 2l^2r^2\cos2\theta)^{1/2}} + 1 - \cos\theta_3 + \cos\theta_1 \right ) d\phi
\end{align}
In the limit $\frac{\Gamma_{12}}{\Gamma_{31}}, \frac{\Gamma_{12}}{\Gamma_{23}} \to 0$, we find that:
\begin{align}
l^2 - r^2 &\sim \rho^2\Gamma_{12}^2\\
(l^4+r^4 - 2l^2r^2\cos2\theta)^{1/2} &\sim \rho^2\Gamma_{31}\Gamma_{12}\\
1-\cos2\theta &\sim \frac{\rho^2\Gamma_{12}^2}{\Gamma_{31}^2}\\
\cos\theta_1 &\sim \frac{\rho^2\Gamma_{12}^2}{\Gamma_{31}^2}\\
1-\cos\theta_3 &\sim \frac{\rho^2\Gamma_{12}^2}{\Gamma_{31}^2}
\end{align}
Thus we find that $\omega_{13}\to 0$ in the limit we are considering. By similar considerations, we find that $\omega_{23} \to 0$ at $\Gamma_1$ in the limit we are considering. So in the asymptotic region $r_a \to 0$, we find that $\omega_j = 0$.

\end{appendix}

\end{document}